\documentclass[12pt]{article}
\usepackage{graphicx}
\usepackage{amsmath}

\newcommand{\nRightarrow}{\Rightarrow\kern-1.2em\hbox{/}\kern.8em} %
\newcommand{\BB}{\hbox{I\kern-.2em\hbox{B}}} 
\newcommand{\DD}{\hbox{I\kern-.2em\hbox{D}}} 
\newcommand{\FF}{\hbox{I\kern-.2em\hbox{F}}} 
\newcommand{\NN}{\hbox{I\kern-.2em\hbox{N}}}  
\newcommand{\ZZ}{{{\rm Z}\kern-.28em{\rm Z}}} 
\newcommand{\RR}{\mathop{{\rm I}\kern-.2em{\rm R}}\nolimits} 
\newcommand{\QQ}{\hbox{l\kern-.36em\hbox{Q}}}  
\newcommand{\CC}{\hbox{I\kern-.58em\hbox{C}}}

\begin{document}
\title{A logical analysis of Stapp's
\\ non-locality theorem}
\author{Giuseppe Nistic\`o\\
{\small Dipartimento
di Matematica, Universit\`a della Calabria, Italy}\\
{\small and}\\
{\small
INFN -- gruppo collegato di Cosenza, Italy}\\
{\small email: gnistico@unical.it} } \maketitle \abstract{
According to an argument proposed by Stapp, Quantum Mechanics
violates the Locality Principle if the 
two hypotheses of {\sl Free
Choices} and {\sl No backward-in-time influence} are assumed to
hold, without the need of introducing hidden variables or criteria
of reality. We develop an approach that endows the new hypotheses
with a logico-mathematical formulation which allow us to perform
an analysis of Stapp's argument, based on ordinary, not
counterfactual, logic. According to our results, the analyzed
argument is not able to conclude that Quantum Mechanics violates
the Locality Principle.}
\section{Introduction}
In Physics the {\it Principle of Locality} can be expressed
through the various conditions it implies. One particular locality
condition is the following one. \vskip.5pc\noindent (L) {\sl
Locality Condition.}\quad {\it In Nature, if\, ${\mathcal
R}_\alpha$ and ${\mathcal R}_\beta$ are two space-time regions
separated space-like, then operations completely confined in
${\mathcal R}_\alpha$, whose realization depends on free choices
made in ${\mathcal R}_\alpha$, must have no influence in
${\mathcal R}_\beta$.} \vskip.5pc An empirical theory is said to
be a {\it local} theory if it is consistent with the principle of
locality. Therefore, all predictions of a local theory must not
contradict the locality condition (L).
\par
Quantum Mechanics, as an empirical theory, establishes
relationships between occurrences of physical events -- including
the occurrences of measurements' outcomes -- if the physical
system is assigned a given state vector $\mid\psi\rangle$; the
quantum theoretical predictions, all confirmed by the experimental
observations so far performed, {\it per se} entail {\it no}
violation of locality conditions. In fact, conflicts between
Quantum Mechanics and locality arise only if further conditions,
which do not belong to the genuine set of quantum postulates, are
required to hold. \par For instance, in order to show inconsistency
between Quantum Mechanics and locality, Bell and his followers
\cite{1}\cite{2}\cite{3}\cite{4} had to introduce the existence of
{\it hidden variables} as further condition, i.e. they needed to
assign pre-existing values to observables which are not measured.
The need for hidden variables is often based, in the literature, on
the criterion or reality \cite{5}:\vskip.35pc \noindent (R) {\sl
Criterion of Reality.}\quad{\it If, without in any way disturbing
a system, we can predict with certainty the value of a physical
quantity, then there exists an element of physical reality
corresponding to this physical quantity}. \vskip.35pc\noindent The
strategy put into effect by these ``classical'' non-locality
theorems for reaching inconsistency between Quantum Mechanics and
locality consists in showing that under certain ideally realizable
circumstances the value assignment entailed by the further
conditions in agreement with quantum mechanics and with the
locality principle is contradictory. \vskip.7pc A different
approach leading to the need for faster than light transfer of
information was followed by Stapp \cite{6}. He found
unsatisfactory the alleged demonstrations of Bell and followers,
just because they ``rest explicitly or implicitly on the
local-hidden-variable assumption that the values of the pertinent
observables exist whether they are measured or not. That
assumption conflicts with the orthodox quantum philosophy''
\cite{7}. Instead of assuming the further conditions entailing a
value assignment to non-measured observables, he supplemented the
standard quantum postulates with two new assumptions:
\begin{description}
\item[{\rm (NBITI)}] one assumption asserts that
{\it once a measurement outcome has actually occurred in a region,
no action in a space-like separated future region can change its
value};
\item[{\rm (FC)}] the other assumption establishes that {\it given a concrete
specimen of the physical system, the choice made in each region as
to which experiment will be performed in that region is a
localized free variable.}
\end{description}
Then he exploited a set of {\sl quantum predictions}, we
collectively refer to as (Q), for a particular physical setting
\cite{3}. This physical setting identifies four quantum
observables $D^{(1)}$, $D^{(2)}$, $B^{(1)}$, $B^{(2)}$ and a state
vector $\vert\psi\rangle$ such that, in particular, the operations
for measuring $D^{(1)}$, $D^{(2)}$ are completely confined in a
space-time region ${\mathcal R}_\alpha$ which is separated
space-like from a region ${\mathcal R}_\beta$ where the operations
for measuring $B^{(1)}$, $B^{(2)}$ are completely confined.
\par
The conditions {(NBITI)}, {(FC)}, and {(Q)} constitute the
hypotheses of Stapp's theorem aiming to show a violation of the
locality condition (L). The proof is hinged on a statement {(SR)},
we make explicit later, having the status of a physical law within
${\mathcal R}_\beta$. From the three hypotheses Stapp proved two
propositions: \vskip.5pc\noindent {\bf Proposition 1.} If
$D^{(2)}$ is measured in ${\mathcal R}_\alpha$ then (SR) holds in
${\mathcal R}_\beta$. \vskip.4pc\noindent {\bf Proposition 2.} If
$D^{(1)}$ is measured in ${\mathcal R}_\alpha$ then (SR) does not
hold in ${\mathcal R}_\beta$. \vskip.5pc\noindent Therefore,
following Stapp, the locality condition (L) is violated because
the {\it validity of law (SR) in ${\mathcal R}_\beta$ turns out to
depend on which measurement is freely chosen to perform in
${\mathcal R}_\alpha$}. This different non-locality proof is
coherent with the starting motivation; indeed, no assignment of
values to unmeasured observables is required. \vskip.5pc Now, the
conditions for the validity of the previously considered classical
non-locality proofs \cite{1}-\cite{4} underwent deep analyses
\cite{8}-\cite{10}; as a result, the possibility of recovering
locality to quantum mechanics turns out to be open. For instance,
in \cite{8},\cite{9} Adenier and Khrennikov show how a bias in the
sample selection involved in Bell's theorem allow for a local
realistic explanation of the violation of Bell's inequality. In
\cite{10} the theorems which adopt the criterion of reality as
further condition, like the theorems of Bell and followers
\cite{1}-\cite{3}, have been analyzed. It has been shown that the
theorems are valid if a {\it wide interpretation} of the criterion
of reality is adopted. But they fail if the criterion is
interpreted according to its {\it strict meaning}; therefore,
these theorems can be re-interpreted as arguments against the wide
interpretation of the criterion of reality rather than as proofs
of non-locality. \vskip.5pc The aim of the present work is to
perform an analysis also of Stapp's non locality theorem, to
establish whether the possibility of recovering locality is open
also in this case. The above explained profound difference between
the logical mechanism of Stapp's proof and that of Bell type
theorems makes ineffective the methodologies used in
\cite{8}-\cite{10}. Our method for the present case consists in
endowing the theoretical apparatus with terms and relations which
enable to express Stapp's new concepts and their consequences as
formal statements, so that Stapp's argument can be reformulated
into a language more suitable for a logical analysis. The results
of such an analysis show that the proof of Proposition 2 drops
into a logical pitfall within Stapp's theoretical scheme developed
by us. Therefore we cannot conclude that a violation of locality
follows from Stapp's argument. \vskip1pc In section 2 Stapp's
argument is described within a suitable theoretical apparatus.
Once established the basic quantum formalism for describing the
concepts at issue, in subsection 2.1 the hypotheses of the theorem
are formulated. The logical structure of the argument is shown in
subsection 2.2, where it is made clear that it yields the
violation of locality if two specific statements, Proposition 1
but also Proposition 2 hold.
\par
Section 3 is devoted to perform the analyses of the proofs of
Proposition 1 and of Proposition 2 from the point of view of
ordinary logic. To do this, in subsection 3.1 we express the
consequences of the new hypotheses introduced by Stapp, namely
(FC) and (NBITI), as {\sl formal statements} within the
theoretical apparatus developed for describing Stapp's approach.
Then the analyses of the proofs of Proposition 1 and Proposition 2
are respectively performed in subsection 3.2 and 3.3 on a
logico-mathematical ground. While Proposition 1 turns out to be
valid, the proof of Proposition 2 drops into an explicitly shown
logical pitfall which invalidates the proof.
\par
In section 4 we show how our work relates to the literature about
the subject.
\section{The theorem}
Given a quantum state vector $\vert\psi\rangle$ of the Hilbert space
$\mathcal H$ which describes the physical system, let ${\mathcal
S}(\vert\psi\rangle)$ be a {\it support} of $\psi$, i.e. a concrete set of
specimens of the physical system whose quantum state is represented by $\vert\psi\rangle$. Let
$D$ be any {\sl two-value} observable, i.e. an observable having only two
possible values denoted by $-1$ and $+1$, and hence represented by a self-adjoint
operator $\hat D$ with purely discrete spectrum
$\sigma(\hat D)=\{-1,+1\}$.
Fixed any support ${\mathcal S}(\vert\psi\rangle)$, every two-value observable $D$ identifies
the following extensions in
${\mathcal S}(\vert\psi\rangle)$:
\begin{description}
\item[-]
the set  ${\bf D}$ of
specimens in ${\mathcal S}(\vert\psi\rangle)$ which {\it actually} undergo a
measurement of $D$;
\item[-]
the set ${\bf D}_+$ of specimens of ${\bf D}$ for which the
outcome $+1$ of $D$ has been obtained;\par
\item[-]
the set ${\bf D}_-$ of specimens of ${\bf D}$ for which the outcome of $D$ is $-1$.
\end{description}
On the basis of the meaning of these concepts
we can assume that the following
statements hold (see \cite{10}, p.1268).
\begin{description}
\item[($2.i$)]
If $D$ is a two-value observable then
for all $\vert\psi\rangle$ a support
${\mathcal S}(\vert\psi\rangle)$
exists  such that ${\bf D}\neq\emptyset$;
\item[($2.ii$)]
${\bf D}_+\cap{\bf D}_-=\emptyset$ and
${\bf D}_+\cup{\bf D}_-={\bf D}$;
\item[($2.iii$)]
If $\langle\psi\vert \hat D\psi\rangle\neq -1$ then ${\mathcal
S}(\vert\psi\rangle)$ exists such that ${\bf D}_+\neq\emptyset$,
and
\item[\qquad\;]
if $\langle\psi\vert \hat D\psi\rangle\neq +1$, then ${\mathcal
S}(\vert\psi\rangle)$ exists such that ${\bf D}_-\neq\emptyset$;
\end{description}
\par
Two observables $D$ and $B$ are {\sl separated}, written $D\bowtie
B$, if their respective measurements require operations confined
in space-like separated regions ${\mathcal R}_\alpha$ and
${\mathcal R}_\beta$. \vskip.4pc According to standard quantum
theory, two observables $D$ and $B$ can be measured together if
and only if the corresponding operators commute with each other;
therefore also the following statements hold for any pair of
two-value observables $D$, $B$.
\begin{description}
\item[($2.iv$)]
$
[\hat D, \hat B]\neq{\bf 0}\quad\hbox{implies}\quad {\bf
D}\cap{\bf B}=\emptyset\hbox{ for all }{\mathcal
S}(\vert\psi\rangle)
$;
\item[($2.v$)]
$
[\hat D, \hat B]={\bf 0}\quad\hbox{ implies }
\quad\forall\psi\;\;\exists{\mathcal S}(\vert\psi\rangle)\hbox{
such that }{\bf D}\cap{\bf B}\neq\emptyset$.
\end{description}
\par
Given a pair $D,B$ of two-value observables such that $[\hat
D,\hat B]={\bf 0}$, we say that the correlation $D\to B$ holds in
the quantum state $\psi$ if, whenever both $A$ and $B$ are
actually measured, i.e. if $x\in{\bf D}\cap{\bf B}$, then
$x\in{\bf D}_+$ implies $x\in{\bf B}_+$; so we have the following
definition. \vskip.5pc\noindent ($3.i$)\quad $D\to B$\quad if
\quad $x\in{\bf D}_+$ implies $x\in{\bf B}_+$, whenever $x\in{\bf
D}\cap{\bf B}$. \vskip.5pc\noindent This correlation admits the
following mathematical characterization \cite{11}.
$$
D\to B\quad\hbox{iff}\quad {{\bf 1}+{\hat D}\over 2}{{\bf 1}+{\hat
B}\over 2}\psi={{\bf 1}+{\hat D}\over 2}\psi. \leqno(3.ii)
$$

\subsection{Hypotheses of the theorem}
Now we explicitly establish the three premises of Stapp's theorem.
\vskip.5pc\noindent {\bf (FC)} {\it Free Choices}: ``This premise
asserts that the choice made in each region as to which experiment
will be performed in that region can be treated as a localized
free variable.''\cite{6} \vskip.5pc\noindent {\bf (NBITI)} {\it No
backward in time influence}: ``This premise asserts that
experimental outcomes that have already occurred in an earlier
region [...] can be considered fixed and settled independently of
which experiment will be chosen and performed later in a region
spacelike separated from the first.''\cite{6} \vskip.7pc\noindent
{\bf (Q)}\quad The {\it third premise} affirms the existence, as
established by Hardy \cite{3} according to quantum mechanics, of
four observables $D^{(1)}$, $D^{(2)}$, $B^{(1)}$, $B^{(2)}$ and of
a particular state vector $\vert\psi\rangle$, for a certain
physical system, which satisfy the following conditions:
\begin{description}
  \item[{\rm (q.i)}] $D^{(1)}$, $D^{(2)}$ are confined in a region ${\mathcal R}_\alpha$
  separated space-like from the region ${\mathcal R}_\beta$ wherein the observables $B^{(1)}$ and $B^{(2)}$
  are confined, with ${\mathcal R}_\alpha$ lying in time earlier than ${\mathcal R}_\beta$.
  Hence in particular $D^{(j)}\bowtie B^{(k)}$, $j,k\in\{1,2\}$.
  \item[{\rm (q.ii)}] $[\hat D^{(1)},\hat D^{(2)}]\neq{\bf 0}$, $[\hat B^{(1)},\hat B^{(2)}]\neq{\bf 0}$;
$-1\neq\langle\psi\vert\hat D^{(j)}\psi\rangle\neq +1$, $-1\neq\langle\psi\vert\hat B^{(j)}\psi\rangle\neq +1$.
  \item[{\rm (q.iii)}]
  $[{\hat D}^{(j)},{\hat B}^{(k)}]={\bf 0}$, $j.k\in\{1,2\}$ and
in the state vector $\vert\psi\rangle$ the following chain of
correlations holds.
  $$a) \; D^{(1)}\to B^{(1)},\quad b)\; B^{(1)}\to D^{(2)},\quad c)\;D^{(2)}\to B^{(2)}.$$
  \item[{\rm (q.iv)}] \quad
  ${\mathcal S}(\vert\psi\rangle)$ and $x_0\in{\mathcal S}
(\vert\psi\rangle)$\quad exist
such that
$x_0\in{\bf D}^{(1)}_+\cap{\bf B}^{(2)}_-$.
\end{description}
In fact, this last condition is implied from the following non-equality satisfied
by Hardy's setting.
   $$\left\langle\psi\mid{{\bf 1}+{\hat D}^{(1)}\over 2}{{\bf 1}-{\hat B}^{(2)}\over 2}\psi\right\rangle\neq 0.\eqno(4)$$
Since the l.h.s. is nothing else but the quantum probability that a simultaneous measurement of
$D^{(1)}$ and $B^{(2)}$ yields respective outcomes $+1$ and $-1$, the non-equality
states that the correlation $D^{(1)}\to B^{(2)}$ does not hold. Therefore, by (3.i) it implies (q.iv).
\subsection{Logical structure of the proof}
The logical mechanism of the non-locality proof at issue is based
on the following pivotal statement. \vskip.35pc\noindent
\begin{description}
\item[(SR)]
``If [$B^{(1))}$] is performed and gives outcome [$+1$], then, if, instead, [$B^{(2)}$]
had been performed the outcome would have been [$+1$].''\cite{6}
\end{description}
The concept of ``instead'' in (SR) can be given a precise,
unambiguous formulation within the theoretical apparatus developed
by supplementing  Quantum Mechanics with the further assumptions
(NBITI) and (FC); this is shown in the next section. Thereofore,
we have to recognize, following Stapp, that (SR) has the status of
a physical law about outcomes of measurements completely
performable within region ${\mathcal R}_\beta$. Then Stapp
introduces the following propositions. \vskip.7pc\noindent {\bf
Propostion 1.} If a measurement of $D^{(2)}$ is performed in
region ${\mathcal R}_\alpha$, then (SR) is valid. \par\noindent
Equivalently,
$$\hbox{Statement (SR) holds in }{\mathcal R}_\beta \hbox{ for all }x\in{\bf D}^{(2)}.$$
{\bf Proposition 2.} If a measurement of $D^{(1)}$ is performed in
region ${\mathcal R}_\alpha$, then (SR) is not valid. Equivalently,
$$\hbox{Statement (SR) does not hold in } {\mathcal R}_\beta \hbox{ for all }x\in{\bf D}^{(1)}.$$
\vskip.5pc\noindent If both these propositions hold, then the
validity of statement {(SR)} in ${\mathcal R}_\beta$ would depend
on what is freely decided to measure in region ${\mathcal
R}_\alpha$, separated space-like from ${\mathcal R}_\beta$; hence
a violation of the locality condition (L) would happen.
\par
In fact, Stapp gives his own proofs \cite{6} that {\it both}
Proposition 1 and Proposition 2 actually follow from the premises
{(FC)}, {(NBITI)} and (Q). Accordingly, the locality condition (L)
should be violated if the three premises hold.
\section{Logical analysis}
In this section we shall examine, from a mere logical point of
view, the proofs of Proposition 1 and Proposition 2 as drawn by
Stapp. Let us begin by considering Proposition 1.
\vskip.5pc\noindent {\bf Proposition 1.}\quad $x\in{\bf
D}^{(2)}\quad\hbox{implies\quad (SR) holds for this }x$.
\vskip.35pc\noindent {\bf Stapp's Proof.}\quad
 ``The concept of `instead' [in {(SR)}] is given a unambiguous meaning
by the combination of the premises of `free' choice and `no
backward in time influence'; the choice between [$B^{(2)}$] and
[$B^{(1)}$] is to be treated, within the theory, as a free
variable, and switching between [$B^{(2)}$] and [$B^{(1)}$] is
required to leave any outcome in the earlier region [${\mathcal
R}_\alpha$] undisturbed. But the statements [(q.iii.a) and
(h.iii.b)] can be joined in tandem to give the result (SR)''
\cite{6}. \vskip.7pc The steps of this proof are carried out by
appealing to their intuitiveness, rather than by means of the
usual logico-mathematical methods; for instance,
how the concept of instead can be given a unambiguous meaning,
and which unambiguous meaning, is not explicitly shown.
In this form the proof
unfits for the aimed analysis. This lack is fulfilled in
the following subsection; we shall endow
the
concept of `instead' with a mathematical counterpart
within the theoretical apparatus, in order to make explicit its
role and formally verifiable the proofs.
\subsection{The meaning of ``instead''}
In order to comply with the starting motivation of Stapp's
approach, the phrase \vskip.4pc
 ``{\it if, instead, $B^{(2)}$ had been performed the
outcome would have been $+1$''} \vskip.4pc\noindent in (SR) must
entail no assignment of a pre-existing value $+1$ to $B^{(2)}$,
since $B^{(2)}$ is a non-measured observable. Then the phrase must
be interpreted as a {\it prediction} about the outcome of a
measurement of $B^{(2)}$, which is valid also if the measurement
of $B^{(2)}$ is not performed. Of course, this kind of validity
goes beyond the conceptual domain of standard Quantum Mechanics;
but our task is just to establish whether it can be consistently
introduced within Stapp's  theoretical framework where {(NBITI)}
and {(FC)} add to standard Quantum Mechanics. Then we introduce
the formal relation
$$
x\in\BB_+^{(2)}
$$
to indicate that the prediction in the phrase is {\it valid} for a
given specimen $x\in{\mathcal S}(\vert\psi\rangle)$. In general,
given a two-value observable $B$, by $\BB_+$ (resp., $\BB_-$) we
shall denote a set of specimens $x\in{\mathcal
S}(\vert\psi\rangle)$ for which the prediction of the outcome $+1$
(resp., $-1$) for a measurement of $B$ is valid, also if the
measurement of $B$ is not actually performed. The consistency of
this new concept requires that the following statements hold.
$$
\BB_+\cap\BB_-=\emptyset;\leqno(5.i)
$$
$$
x\in\BB_-\hbox{ implies } x\notin{\bf B}_+\quad\hbox{and}\quad
x\in\BB_+\hbox{ implies } x\notin{\bf B}_-.\leqno(5.ii)
$$
As above noticed, statements such as $x\in\BB_+$ or $x\in\BB_-$ do
not make sense within standard Quantum Mechanics; our next step is
to single out {\it which conditions} make valid these kind of
statements in a theory where the postulates of standard Quantum
Mechanics are supplemented with the assumptions {(FC)} and
{(NBITI)} of Stapp's approach.
\par
Let $D$ be a two-value observable confined in a region ${\mathcal
R}_\alpha$ separated space-like from another region ${\mathcal
R}_\beta$ where two other two-value observables $B$ and $F$ are
confined, with ${\mathcal R}_\alpha$ located in time before
${\mathcal R}_\beta$; moreover, let the empirical implications
$D\to B$ and $D\to F$ hold, with $[\hat B,\hat F]\neq{\bf 0}$ and
hence ${\bf B}\cap{\bf F}=\emptyset$ according to (2.iv).
Let us suppose that in
${\mathcal R}_\alpha$ the observable $D$ is actually measured on a
specimen $x\in{\mathcal S}(\vert\psi\rangle)$ and that the outcome
$+1$ is obtained; i.e., we are supposing that $x\in{\bf D}_+$.
Now, if later in ${\mathcal R}_\beta$ the observable $F$ is chosen
to be measured on this $x$, i.e. if $x\in{\bf F}$, then the
prediction, made before measuring $F$, that the outcome will be
$+1$ is valid according to standard Quantum Mechanics because the
following conditions hold.
\begin{description}
\item[{\rm (C.1)}] The
outcome of $D$ is not changed by the choice of measuring $F$, by
{(NBITI)}.
\item[{\rm (C.2)}] Such an outcome is $+1$.
\item[{\rm (C.3)}]
$D\to F$ and $x\in{\bf D}\cap{\bf F}$, so that (3.i) applies.
\end{description}
But if we assume that also {(FC)} holds, then the choice of
measuring $F$ or $B$ (or another observable) in ${\mathcal
R}_\beta$ is {\it free}, so that at this stage, before choosing
the measurement to perform in ${\mathcal R}_\beta$, $F$ and $B$
are on the same footing in our theory. Then the prediction that
the outcome of a measurement of $B$ will be $+1$ must be equally
valid with respect to the prediction for $F$, also if $B$ is not
measured because ${\bf B}\cap{\bf F}=\emptyset$; hence in this
case (FC) and (NBITI) imply that the relation $x\in\BB_+$ is
valid. Thus we can conclude, in general, that the following
statement holds\footnote{The validity of conclusion (5.iii) seems
to depend on the existence of an observable $F$ satisfying the
conditions required by the argument. In fact, such an existence is
ensured under quite general conditions. For instance, if such an
observable $F$ does not exist in ${\mathcal R}_\beta$, it will
exist in another region ${\mathcal R}_{\beta'}$ suitably shifted
in space-time with respect to ${\mathcal R}_\beta$, so that the
region ${\mathcal R}_\gamma={\mathcal R}_\beta\cup{\mathcal
R}_{\beta'}$ is still space-like separated from ${\mathcal
R}_{\alpha}$ and in its future. Therefore our reasoning carried
out with ${\mathcal R}_{\gamma}$ replacing ${\mathcal R}_{\beta}$
will lead to the conclusion (5.iii).}.
$$
D\to B\quad\hbox{implies}\quad x\in{\bf D}_+\Rightarrow x\in{\BB}_+\leqno(5.iii)
$$
{\it if $D$ and $B$ are two-value observables respectively confined
in space-like separated regions ${\mathcal R}_\alpha$ and ${\mathcal R}_\beta$,
with ${\mathcal R}_\alpha$ located in time before ${\mathcal R}_\beta$.}
\vskip.7pc\noindent
{\bf Remark 3.1.}\quad
It must be stressed that the stronger implication
$$
\hbox{ If }\quad D\to B\quad\hbox{then}\quad
x\in\DD_+\Rightarrow x\in{\BB}_+\leqno(5.iv)
$$
cannot be inferred under the same conditions for $D$ and $B$
by extending the argument which makes use of
(NBITI) and (FC) for inferring (5.iii). Indeed, in (5.iv) no {\it
actual} outcome for $D$ is available which enable us to invoke
Quantum Mechanics for making a prediction of the outcome for $F$
or $B$. Moreover, the absence of the actual outcome forbids to
apply {(NBITI)}.
\vskip.5pc Hence, the supplementation of
quantum mechanics with the assumptions {(NBITI)} and {(FC)}
entails conditions (5.i)-(5.iii) which give the concept of
``instead'' an unambiguous consistent formulation. The new
theoretical concepts make possible to re-formulate the crucial
statement {(SR)} of Stapp's argument in the following neologized
simple form
$$
x\in{\bf B}^{(1)}_+\quad\hbox{implies}\quad
x\in{\BB}^{(2)}_+,\leqno{(SR)}_\nu
$$
with evident gain in formal precision and clarity.
\subsection{Proposition 1.}
Now we can perform the analysis of Proposition 1's proof.
\vskip.4pc\noindent {\bf Proposition 1.}\quad $x\in{\bf
D}^{(2)}\Rightarrow$ ${(SR)}_\nu$ holds in ${\mathcal
R}_\beta$. \vskip.4pc\noindent {\bf Logical expansion of the
proof.}
\begin{description}
\item[{\rm (E.1)}] Let us suppose that the antecedent of {Proposition 1} holds:
$$x\in{\bf D}^{(2)}.\eqno(6.i)$$
\item[{\rm (E.2)}]
Let us suppose that the antecendent of ${(SR)}_\nu$ holds too:
$$x\in{\bf B}^{(1)}_+.\eqno(6.ii)$$
\item[{\rm (E.3)}]
Hence (6.i) and (6.ii) imply
$$x\in {\bf B}^{(1)}\cap{\bf D}^{(2)}.\eqno(6.iii)$$
\item[{\rm (E.4)}] Then
(q.iii.b), (6.ii) and (6.iii) imply
$$x\in {\bf D}^{(2)}_+.\eqno(6.iv)$$
\item[{\rm (E.5)}]
(q.iii.c), (6.iv) and (5.iii) imply
$$x\in{\BB}^{(2)}_+.$$
\end{description}
\vskip.35pc\noindent Since the validity of the steps from (E.3) to
(E.5) is self-evident, in order that this re-worded proof be
completely correct, it is sufficient to prove that specimens
satisfying (6.i) and (6.ii) actually exist. Now, by (q.iii.b),
(3.ii) and (q.ii) we have ${{\bf 1}+{\hat B}^{(1)}\over 2}{{\bf
1}+{\hat D}^{(2)}\over 2}\psi={{\bf 1}+{\hat B}^{(1)}\over
2}\psi\neq 0$. Therefore $\langle\psi\mid {{\bf 1}+{\hat
B}^{(1)}\over 2}{{\bf 1}+{\hat D}^{(2)}\over 2}\psi\rangle\neq 0$.
But this last is just the probability that a simultaneous
measurement of $B^{(1)}$ and $D^{(2)}$ yields respective outcomes
$+1$ and $+1$; being it non vanishing we have to conclude that
specimens $x$ satisfying (6.i) and (6.ii) actually exist.
\vskip.35pc Thus our analysis does agree with Stapp's conclusion
that (SR) holds if $D^{(2)}$ is measured in ${\mathcal R}_\alpha$.
\subsection{Proposition 2.}
Now we submit Proposition 2 to our analysis. \vskip.5pc\noindent
{\bf Proposition 2.}\quad $x\in{\bf D}^{(1)}\nRightarrow$ (SR)
holds for this $x$ in ${\mathcal R}_\beta$. \vskip.4pc\noindent
This Proposition is equivalent to the following statement.
\vskip.4pc \centerline{\it $x_0\in {\bf D}^{(1)}$ exists such that
the antecedent of (SR) is true but the consequent is false}
\vskip.4pc\noindent By making use of the neologized form
{(SR)}$_\nu$, such a statement admits the following formulation.
  $$\exists x_0\in{\bf D}^{(1)},\quad x_0\in{\bf B}^{(1)}_+\quad\hbox{but}\quad x_0\notin{\BB}^{(2)}_+.\eqno(7)$$
\vskip.35pc\noindent {\bf Stapp's Proof}\,:\quad ``Quantum theory
predicts that if [$D^{(1)}$] is performed, then outcome [+1]
appears about half the time. Thus, if [$D^{(1)}$] is chosen, then
there are cases where [$x\in{\bf D}^{(1)}_+$] is true. But in a
case where [$x\in{\bf D}^{(1)}_+$] is true, the prediction
[(q.iii.a)] asserts that the premise of (SR) is true. But
statement [(q.iv)], in conjunction with our two premises that give
meaning to `instead', implies that the conclusion of (SR) is not
true: if [$B^{(2)}$] is performed instead of [$B^{(1)}$], the
outcome is not necessarily [+1], as it was in case [$D^{(2)}$].
So, there are cases where [$D^{(1)}$] is true but (SR) is false.''
\cite{6} \vskip.5pc\noindent Conclusion (7) is attained by Stapp
through the following sequence of statements we translate from his
proof. \vskip.5pc\noindent
\begin{description}
\item[{\rm (S.1)}]
A support ${\mathcal S}(\vert\psi\rangle)$ exists such that ${\bf
D}_+^{(1)}\neq\emptyset$.
\item[{\rm (S.2)}]
$x\in{\bf D}_+^{(1)}\Rightarrow x\in{\bf B}_+^{(1)}$.
\item[{\rm (S.3)}]
The antecedent of ${(SR)}_\nu$ holds $\forall x\in{\bf
D}_+^{(1)}$.
\item[{\rm (S.4)}]
$\exists x_0\in{\bf D}_+^{(1)}$ such that $x_0\in{\bf B}_-^{(2)}$.
\item[{\rm (S.5)}]
$x_0\notin\BB^{(2)}_+$.
\end{description}
\vskip.5pc\noindent Let us now check the validity of each step.
\vskip.5pc\noindent Statement (S.1) holds by (2.iii) and (q.ii).
\vskip.5pc\noindent Statement (S.3) is implied from (S.1) and
(S.2). \vskip.5pc\noindent Statement (S.4) holds because of
(q.iv). \vskip.5pc\noindent Statement (S.5) holds because of (S.4)
and (5.ii). \vskip.7pc\noindent We see that all steps (S.1),
(S.3), (S.4), (S.5) hold true according to a logical analysis.
\vskip.5pc But statement (S.2), is it valid? Statement (S.2) is
nothing else but the translation into our language of the phrase
``But in a case where [$x\in{\bf D}^{(1)}_+$] is true, the
prediction [(q.iii.a)] asserts that the premise of (SR) is true''
stated by Stapp in his proof. Hence, according to Stapp's proof,
(S.2) holds because of (q.iii.a) $D^{(1)}\to B^{(1)}$. But
$$x\in{\bf D}^{(1)}_+\Rightarrow x\in{\bf B}^{(1)}_+
$$
follows from $D^{(1)}\to B^{(1)}$ {\it if $x\in{\bf
D}^{(1)}\cap{\bf B}^{(1)}$ holds too}, because of (3.i). However,
this last condition cannot hold for the specimen $x_0$ considered
in (S4), because it has been characterized by the two conditions
$x_0\in{\bf D}^{(1)}_+$ and  $x_0\in{\bf B}^{(2)}_-$. But if
$x_0\in{\bf B}^{(2)}_-$ holds then $x_0\in{\bf B}^{(2)}$ obviously
holds too, so that the premise of (SR), $x_0\in{\bf B}^{(1)}_+$,
cannot hold because $B^{(1)}$ and $B^{(2)}$ do not commute with
each other and therefore ${\bf B}^{(1)}\cap{\bf
B}^{(2)}=\emptyset$, by (q.ii) and (2.iv). \vskip.4pc\noindent
{\bf Remark 3.2.}\quad In fact, a contradiction would arise if
(5.iv) in remark 3.1 were valid. Indeed, the statement
$$x\in{\bf D}^{(1)}_+\Rightarrow x\in\BB_+^{(1)}\eqno(8)$$ can be deduced from
the relation $D^{(1)}\to B^{(1)}$ stated by the prediction
(q.iii.a) and by (5.iii). From (8), (q.iii.b) and (5.iv) we could
deduce $x\in\DD^{(2)}_+$; thereby $x\in\BB_+^{(2)}$ would follow,
by (q.iii.c) and (5.iv). But $x\in\BB^{(2)}_+$ would imply
$x\notin{\bf B}^{(2)}_-$, by (5.ii). Therefore, if (5.iv) were valid,
the statement
$$
x\notin{\bf B}^{(2)}_-\quad\hbox{for all}\quad x\in{\bf D}^{(1)}_+
$$
would be true. This statement contradicts quantum mechanics
prediction (q.iv). However, such a contradiction  would be not a
proof of Proposition 2, because it does not concern with the
validity of ${(SR)}_\nu$. In other words, Proposition 2 is not
even proved by assuming (5.iv). Rather, this contradiction would
prove that Quantum Mechanics is not consistent with {(NBITI)} and
{(FC)}, if (5.iv) could be inferred from {(NBITI)} and {(FC)}; but
this is not the case, as stressed in remark 3.1.
\section{Final comments}
The object of the analysis performed in the present work is the
final result of an investigation pursued by Stapp during many
years \cite{12}-\cite{18} with the aim of proving inconsistency
between Quantum Mechanics and locality without the need of
attributing values to unmeasured observables. The earlier
proposals \cite{19}-\cite{21} were explicitly based on {\it
counterfactual reasonings}; they received severe criticisms
\cite{22},\cite{23} which questioned the validity of the proof
just on the ground of {\it counterfactuals theory} \cite{24}.
Stapp disputed \cite{21},\cite{25} these criticisms, but also
expressed dissatisfaction with proofs based on counterfactual
concepts: ``[...] these theories, though useful in other ways, do
not provide a completely adequate foundation for the study of the
deep physical question of locality: basic physical conclusions
should not rest on arbitrary conventions [which infiltrate
counterfactuals theory]'' \cite{25}. Then he presented
improvements of the proof, with the aim of making the argument
valid without the need of rules of counterfactuals theory
\cite{26},\cite{27},\cite{16},\cite{17}, until the final version
\cite{6} about which he states: ``my 2004 proof, although
retaining some of the trappings and language of counterfactual
argumentation, is based on a substantially different foundation.
The combination of my assumptions of `free choices' and of 'no
backward-in-time influence' amounts to the assumption that
theories covered by my new work {\it are to be compatible with}
the idea of `fixed past, open future'. This conceptualization
circumvents, at the foundational level, the need for
counterfactuals'' \cite{7}.
\par
However the debate has not reached a shared conclusion, because
the criticisms about the counterfactual character of the proof
continued \cite{28}-\cite{31}. On the other hand, Stapp always
answered by essentially arguing that the new formulations of the
proof do not make use of counterfactuals theory
\cite{18},\cite{7}.
\par
The analysis performed in the present work differs from the
previous criticisms. The key step of our work is realized in
subsection 3.1. Here the theoretical apparatus of Quantum
Mechanics has been endowed with the further statements and
relations (5.i)-(5.iii) which formally express the new assumptions
(FC) and (NBITI) introduced by Stapp. This is done without
invoking counterfactual concepts. In particular, statement (5.iii)
is the translation of the `fixed past, open future' principle into
an appropriate formulation within a coherent theoretical apparatus
developed from Quantum Mechanics. Therefore our investigation is
not based on a criticism of Stapp's methods or on a rejection of
his new concepts. Rather, we develop a coherent theoretical
formulation from his conceptualization, where the proof can be
formulated in  logico mathematical language and its validity
verified by a purely logico-mathematical analysis which shows that
according to our approach the proof is not valid.

\end{document}